\documentclass[superscriptaddress,secnumarabic,amssymb,amsmath,nobibnotes,aps,prd,showkeys,showpacs,nofootinbib,preprint]{revtex4-1}%

\usepackage{graphicx}
\usepackage{bm}
\usepackage{amsmath}
\usepackage{amsfonts}
\usepackage{amssymb}
\usepackage{epstopdf}
\usepackage{natbib}%
\newcommand{\bee}{\begin{equation}}
\newcommand{\eee}{\end{equation}}
\newcommand{\eaa}{\end{eqnarray}}
\newcommand{\baa}{\begin{eqnarray}}
\def\ni{\noindent}

\usepackage{color}

\begin{document}

\title{Tsallis and Kaniadakis statistics from the viewpoint of entropic gravity formalism}

\author{Everton M. C. Abreu}\email{evertonabreu@ufrrj.br}
\affiliation{Grupo de F\' isica Te\'orica e Matem\'atica F\' isica, Departamento de F\'{i}sica, Universidade Federal Rural do Rio de Janeiro, 23890-971, Serop\'edica, RJ, Brazil}
\affiliation{Departamento de F\'{i}sica, Universidade Federal de Juiz de Fora, 36036-330, Juiz de Fora, MG, Brazil}
\author{Jorge Ananias Neto}\email{jorge@fisica.ufjf.br}
\affiliation{Departamento de F\'{i}sica, Universidade Federal de Juiz de Fora, 36036-330, Juiz de Fora, MG, Brazil}
\author{Ed\'esio M. Barboza Jr.} \email{edesiobarboza@uern.br}
\affiliation{Departamento de F\'isica, Universidade do Estado do Rio Grande do Norte, 59610-210, Mossor\'o, RN, Brazil}
\author{Rafael C. Nunes}\email{rafadcnunes@gmail.com}
\affiliation{Departamento de F\'{i}sica, Universidade Federal de Juiz de Fora, 36036-330, Juiz de Fora, MG, Brazil}


\begin{abstract}
\noindent It has been shown in the literature that
effective gravitational constants, which are derived from Verlinde's formalism, can be used to introduce the Tsallis and Kaniadakis statistics. This method provides a simple alternative to the usual procedure normally used in these non-Gaussian statistics. We have applied our formalism in the Jeans mass criterion of stability and in the free fall time collapsing of a self-gravitating system where new results are obtained. A possible connection between our formalism and deviations of Newton's law of gravitation in a submillimeter range is made.
\end{abstract}

\maketitle

An meaningful extension of Boltzman-Gibbs's (BG) statistical theory has been suggested by Tsallis  \cite{tsallis}. This model is also currently referred in the literature as a nonextensive (NE) statistical mechanics. Tsallis thermostatistics formalism defines a nonadditive entropy as
\begin{eqnarray}
\label{nes}
S_q =  k_B \, \frac{1 - \sum_{i=1}^W p_i^q}{q-1}\;\;\;\;\;\;\qquad \Big(\,\sum_{i=1}^W p_i = 1\,\Big)\,\,,
\end{eqnarray}

\ni where $p_i$ is the probability of a system to exist within a microstate, $W$ is the total number of configurations and 
$q$, known in the current literature as being the Tsallis parameter or NE  parameter, is a real parameter which measures the degree of nonextensivity. 
The definition of entropy in Tsallis statistics carries the standard properties of positivity, equiprobability, concavity and irreversibility. This approach has been successfully used in many different physical system. For instance, we can mention the Levy-type anomalous diffusion \cite{levy}, turbulence in a pure-electron plasma \cite{turb} and gravitational systems \cite{sys}.
It is noteworthy to affirm that Tsallis thermostatistics formalism has the BG statistics as a particular case in the limit $ q \rightarrow 1$ where the standard additivity of entropy can be recovered. 

Plastino and Lima, in  \cite{PL}, have obtained a NE equipartition law of energy. It has been demonstrated that the kinetic foundations of Tsallis' NE statistics drives us to a velocity distribution for free particles given by   \cite{SPL}
\begin{eqnarray}
\label{vd}
f_q(v)=B_q \Big[1-(1-q) \frac{m v^2}{2 k_B T}\Big]^{1/1-q},
\end{eqnarray}

\ni where $B_q$ is a normalization constant.   Hence, the expectation value of $v^2$ for each degree of freedom, is given by  \cite{SA}
\begin{eqnarray}
\label{v2}
<v^2>_q=\frac{\int^\infty_0 f_q\,v^2\,  dv}{\int^\infty_0 f_q \, dv}\\ \nonumber
=\frac{2}{5-3q}\, \frac{k_B T}{m}.
\end{eqnarray}

\noindent So, the equipartition theorem can be obtained by using that
\begin{eqnarray}
\label{reqq}
E_q=\frac{1}{2} N m <v^2>_q,
\end{eqnarray}

\ni and we will arrive at
\begin{eqnarray}
\label{ge}
E_q = \frac{1}{5 - 3 q} N k_B T\,\,.
\end{eqnarray}

\ni It is well established that the range of $q$ is $ 0 \le q < 5/3 $.  For $ q=5/3$,  Eq. (\ref{ge}), diverges. 
It is also easy to notice that for $ q = 1$,  the classical equipartition theorem for each microscopic degrees of freedom (dof) can be reacquired.



The well known Kaniadakis statistics  \cite{kani1}, also refereed as $\kappa$-statistics, analogously to Tsallis thermostatistics model, generalizes the usual BG statistics initially by introducing both the $\kappa$-exponential and $\kappa$-logarithm defined respectively by
\begin{eqnarray}
\label{expk}
exp_\kappa(f)=\Big( \sqrt{1+\kappa^2 f^2}+\kappa f \Big)^\frac{1}{\kappa},
\end{eqnarray}
\begin{eqnarray}
\label{logk}
\ln_\kappa(f)=\frac{f^\kappa-f^{-\kappa}}{2\kappa},
\end{eqnarray}

\ni and the following property can be satisfied, namely,
\begin{eqnarray}
\ln_\kappa\Big(exp_\kappa(f)\Big)=exp_\kappa\Big(\ln_\kappa(f)\Big)\equiv f.
\end{eqnarray}

\ni From Eqs. (\ref{expk}) and (\ref{logk}) we can notice that the $\kappa$-parameter twists the standard definitions of the exponential and logarithm functions.

The $\kappa$-entropy connected to this $\kappa$-framework can be written as
\begin{eqnarray}
S_\kappa=- k_B \sum_i^W  \,\frac{p_i^{1+\kappa}-p_i^{1-\kappa}}{2\kappa},
\end{eqnarray}

\ni which recovers the BG entropy in the limit $\kappa \rightarrow 0$. It is relevant to comment here that the $\kappa$-entropy has satisfied the properties concerning concavity, additivity and extensivity. The $\kappa$-statistics has thrived when applied in many experimental scenarios. As an example we can cite cosmic rays  \cite{Kanisca1} and cosmic effects  \cite{aabn-1}, quark-gluon plasma  \cite{Tewe}, kinetic models describing a gas of interacting atoms and photons  \cite{Ross} and financial models  \cite{RBJ}.

The kinetic foundations for the $\kappa$-statistics lead us to a velocity distribution for free particles given by  \cite{bss}
\begin{eqnarray}
f_\kappa(v)=\Big[ \sqrt{1+\kappa^2 \Big( -\frac{m v^2}{2 k_B T}\Big)^2}  - \frac{\kappa m v^2}{2 k_B T} \Big]^\frac{1}{\kappa}.
\end{eqnarray}

\ni Hence, the expectation value of $v^2$, for each dof, is
\begin{eqnarray}
\label{vk}
<v^2>_\kappa=\frac{\int^\infty_0\, f_\kappa \,v^2 dv}{\int^\infty_0\, f_\kappa dv}.
\end{eqnarray}

\ni Using the integral relation  \cite{kani1} we have that
\begin{eqnarray}
\int_0^\infty x^{r-1}\, exp_\kappa (-x) dx
=\,\frac{|2\kappa|^{-r}}{1+r|\kappa|}\; \frac{\Gamma\Big(\frac{1}{\left|2\kappa\right|}-\frac{r}{2}\Big)}{\Gamma\Big
(\frac{1}{\left|2\kappa\right|}+\frac{r}{2}\Big)}\; \Gamma(r)\,\,, \nonumber \\
\end{eqnarray}

\ni which drives us to
\begin{eqnarray}
\label{vk2}
<v^2>_\kappa=\frac{(1+\frac{\kappa}{2})}{(1+\frac{3}{2}\kappa) \,2\kappa}
\frac{\Gamma{(\frac{1}{2\kappa}-\frac{3}{4})\,\Gamma{(\frac{1}{2\kappa}+\frac{1}{4})}}}{\Gamma{(\frac{1}{2\kappa}+\frac{3}{4})\,\Gamma{(\frac{1}{2\kappa}-\frac{1}{4})}}}\;  \frac{k_B T}{m}\,\,.  \nonumber \\
\end{eqnarray}

\ni The $\kappa$-equipartition theorem can be calculated through the relation
\begin{eqnarray}
\label{reqk}
E_\kappa=\frac{1}{2} N m <v^2>_\kappa\,,
\end{eqnarray}

\ni and we arrive at
\begin{eqnarray}
\label{keq}
E_\kappa=\frac{1}{2} N \;\; \frac{(1+\frac{\kappa}{2})}{(1+\frac{3}{2}\kappa) \,2\kappa}
\frac{\Gamma{(\frac{1}{2\kappa}-\frac{3}{4})\,\Gamma{(\frac{1}{2\kappa}+\frac{1}{4})}}}{\Gamma{(\frac{1}{2\kappa}+\frac{3}{4})\,\Gamma{(\frac{1}{2\kappa}-\frac{1}{4})}}}\;\;\; k_B T\,\,, \nonumber \\
\end{eqnarray}

\ni and the range of $\kappa$ is $ 0 \le \kappa < 2/3 $.  For $ \kappa=2/3$ (critical value) the expression for the equipartition law of energy, Eq. (\ref{keq}), diverges, which means that $q=5/3$ in Tsallis formalism. For $ \kappa = 0$, the classical equipartition theorem for each microscopic degrees of freedom can be recovered. 

The recent formalism idealized by E. Verlinde   \cite{verlinde} computes the gravitational acceleration  by using the holographic principle and the well known equipartition law of energy. His ideas are based on the fact that gravitation can be considered universal and independent of the details of the spacetime microstructure.  Besides, he introduced new ideas concerning holography since the holographic principle must unify matter, gravity and quantum mechanics.


The model discuss a spherical surface playing the role of an holographic screen, with a particle of mass $M$ located in its center. The holographic screen can be envisaged as a storage device for information. The number of bits (the underlying unit of information in the holographic screen) is assumed to be proportional to the  holographic screen area $A$ and is represented by
\begin{eqnarray}
\label{bits}
N = \frac{A }{\ell_P^2},
\end{eqnarray}

\ni where $ A = 4 \pi r^2 $ and $\ell_P$ is the Planck length. In Verlinde's framework one can assume that the total number of bits for the energy on the screen is given by the equipartition law of energy
\begin{eqnarray}
\label{eq}
E = \frac{1}{2}\,N k_B T.
\end{eqnarray}
It is important to understand that the usual equipartition theorem in this last equation can be obtained from the standard BG thermostatistics.  Let us deem that the energy of the particle inside the holographic screen is equally divided by all bits in such a way that we can have the equation
\begin{eqnarray}
\label{meq}
M c^2 = \frac{1}{2}\,N k_B T,
\end{eqnarray}

\ni and to calculate the gravitational acceleration, we can use both the Eq. (\ref{bits}) and the Unruh temperature equation   \cite{unruh} given by
\begin{eqnarray}
\label{un}
k_B T = \frac{1}{2\pi}\, \frac{\hbar a}{c}\,\,.
\end{eqnarray}

\ni So, one can be  able to compute the  (absolute) gravitational acceleration formula
\begin{eqnarray}
\label{acc}
a &=&  \frac{l_P^2 c^3}{\hbar} \, \frac{ M}{r^2}\nonumber\\ 
&=& G \, \frac{ M}{r^2}\,\,.
\end{eqnarray}

\ni From Eq. (\ref{acc}) we can see that the Newton constant $G$ can be written in terms of the standard constants

\begin{eqnarray}
\label{glp}
G=\frac{\ell_P^2 c^3}{\hbar}.
\end{eqnarray}

Using the NE equipartition theorem in Verlinde's formalism one can apply the NE equipartition equation, namely, Eq. (\ref{ge}).  In this way we can obtain a modified acceleration formula   \cite{abreu}
\begin{eqnarray}
\label{accm}
a = G_q \, \frac{ M}{r^2},
\end{eqnarray}

\ni where $G_q$ is an effective gravitational constant which can be given by
\begin{eqnarray}
\label{S}
G_q=\,\frac{5-3q}{2}\,G\,\,.
\end{eqnarray}

\ni From this last equation, we can note that the effective gravitational constant depends on the NE parameter $q$. For instance, when $q=1$ we have that $G_q=G$ (BG scenario) and for $q\,=\,5 / 3$ we have the curious and hypothetical result 
$G_q=0$.  

If we use Kaniadakis' equipartition theorem, Eq.(\ref{keq}), in Verlinde's considerations, the modified acceleration equation can be given by
\begin{eqnarray}
\label{acck}
a = G_{\kappa} \, \frac{ M}{r^2}\,\,,
\end{eqnarray}
\ni where $G_{\kappa}$ is an effective gravitational constant which can be written as
\begin{eqnarray}
\label{Gk}
G_{\kappa}=  \frac{(1+\frac{3}{2}\kappa) \,2\kappa}{(1+\frac{\kappa}{2})}\; \;
\frac{\Gamma{(\frac{1}{2\kappa}+\frac{3}{4})\,\Gamma{(\frac{1}{2\kappa}-\frac{1}{4})}}}{\Gamma{(\frac{1}{2\kappa}-\frac{3}{4})\,\Gamma{(\frac{1}{2\kappa}+\frac{1}{4})}} }\; \; G \,\,.
\end{eqnarray}

\ni It can be easily shown that the limits $\kappa=0, q=1$, with which we can recover the BG statistics, and $\kappa=2/3, q=5/3$.  It is important to say that these values have came from the well known   \cite{bss,kaniqk1,kaniqk2} linear connection between $\kappa$ and $q$ represented by
\begin{eqnarray}
\label{rel}
\kappa=q-1\,\,.
\end{eqnarray}

\ni Specifically, the limits $\kappa=0$, when $q=1$, and $\kappa=\frac{2}{3}$, when $q=\frac{5}{3}$, lead to $G_{q}=G_{\kappa}$. It is clear that the relation $(\ref{rel})$ is true only for the range $1\leq q < 5/3$, given by Eq. (13).

Here we will analyze the self-gravitating system where the internal dynamical stability usually can be depicted by the Jeans gravitational instability criterion. In a recent work\cite{nos} we have studied the Jeans length in the context of Kaniadakis' statistics. Initially, we will apply our formalism in the Jeans mass criterion of stability \cite{j1,jeansbook}, which is given by\begin{eqnarray}
\label{mj}
M_J= \left(\frac{5 k_B T}{G m}\right)^{\frac{3}{2}} \, \left(\frac{3}{4\pi\rho}\right)^{\frac{1}{2}}\,\,,
\end{eqnarray}

\ni where $k_B$ is the Boltzmann constant, $T$ is the temperature,  $m$ is the particle mass, $G$ is the gravitational constant and $\rho$ is the equilibrium mass density. The critical value $M_J$ in Eq. (\ref{mj}), is obtained through the consideration of both the virial theorem and the conservation of energy. The Jeans mass means that, if the mass of a self-gravitating system is greater than $M_J$ consequently the system will become gravitationally unstable and, as a result, it will collapse.

Substituting $G$ by $G_q$, Eq. (\ref{S}), into Eq. (\ref{mj}), we can write the expression for the Jeans mass Tsallis' statistics
\begin{eqnarray}
\label{lbdaq}
M_J^q&=&\left(\frac{2}{5-3q}\, \frac{5 k_B T}{Gm}\right)^{\frac{3}{2}} \,\left(\frac{3}{4\pi\rho}\right)^{\frac{1}{2}}\nonumber\\
&=&\left(\frac{2}{5-3q}\right)^{\frac{3}{2}}\, M_J \,\,.
\end{eqnarray}

\ni This NE modification of the Jeans criterion leads us to a new Jeans mass $M_j^q$ that relies on the nonextensive $q$-parameter as the following conditions:
\vskip .2cm
\noindent (i) if $q=1 \Rightarrow M>M_J^q=M_J$, the usual Jeans criterion is maintained;
\vskip .1 cm
\noindent (ii) if $0<q<1$ the Jeans criterion will be modified as $M>M_J^q<M_J$;
\vskip .1 cm
\noindent (iii) if $1<q<5/3$,  the Jeans criterion will be modified as $M>M_J^q>M_J$;
\vskip .1 cm
\noindent (iv) if $q\rightarrow 5/3, M_J^q \rightarrow \infty$, the self-gravitating system is always stable, \\

\ni where we have used both Verlinde's formalism and Tsallis' thermostatistics to obtain NE Jeans' mass. 

Using Eq. (\ref{Gk}) in (\ref{mj}), we can derive the expression for the critical mass in Kaniadakis' statistics
\begin{eqnarray}
\label{lbdak}
M_J^\kappa=\left(\frac{(1+\frac{\kappa}{2})}{(1+\frac{3}{2}\kappa) \,2\kappa}\; 
	\frac{\Gamma{(\frac{1}{2\kappa}-\frac{3}{4})\,\Gamma{(\frac{1}{2\kappa}+\frac{1}{4})}}}{\Gamma{(\frac{1}{2\kappa}+\frac{3}{4})\,\Gamma{(\frac{1}{2\kappa}-\frac{1}{4})}}}\right)^\frac{3}{2} M_J.
\end{eqnarray}

\ni This $\kappa$ modification of the Jeans criterion leads to a new Jeans mass $M_j^\kappa$ that depends on the $\kappa$-parameter as follows:
\vskip .2cm
\noindent (i) if $\kappa=0 \Rightarrow M>M^\kappa=M_J$, the standard Jeans criterion is recovered.
\vskip .1 cm
\noindent (ii) if $0<\kappa<2/3$,  the Jeans criterion is modified as $M>M_J^\kappa>M_J$.
\vskip .1 cm
\noindent (iii) f $\kappa\rightarrow 2/3-, M_J^\kappa\rightarrow \infty$, the self-gravitating system is always stable.

\bigskip \bigskip

In Figure 1 we have plotted $M_J^q$, Eq. (\ref{lbdaq}), and $M_J^\kappa$, Eq. (\ref{lbdak}).  These last mentioned equations were normalized in the curve in Figure 1 by $M_J$, Eq.(\ref{mj}), as function of $\kappa$.  To accomplish it, we have used Eq. (\ref{rel}).
\begin{figure}
\begin{center}
\label{mass_q}
\includegraphics[width=3.in, height=3.in]{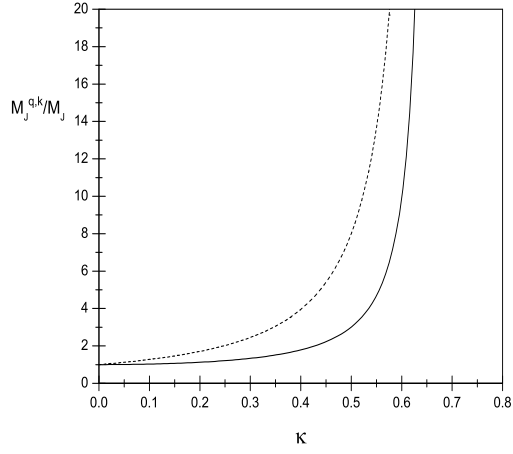}
\caption{Dashed line: Tsallis critical mass, $M_J^q$. Solid line: Kaniadakis critical 
mass, $M_J^\kappa$.}
\end{center}
\end{figure}

From Figure 1 we can notice that, except for the limiting values, $\kappa=0$ and $\kappa=2/3$, the Tsallis critical mass is greater than the Kaniadakis critical mass. This result shows that the NE effects of Tsallis' statistics leads us to a self-gravitating system being more stable when compared with the results produced by Kaniadakis' statistics since $M_J^q > M_J^\kappa$.

Another important physical quantity in the self-gravitating system is the free fall time (FFT) which expression is given by

\begin{eqnarray}
\label{ff}
t_{ff}=\sqrt{\frac{3}{2\pi G \rho}}.
\end{eqnarray}

\ni The FFT is the characteristic time that would take to the cloud or the self-gravitating system to finally collapse.
Substituting $G$ by $G_q$, Eq.(\ref{S}), in (\ref{ff}), we can obtain the expression for the free fall time in the Tsallis statistics

\begin{eqnarray}
\label{ffq}
t^q_{ff}=\left(\frac{2}{5-3q}\right)^\frac{1}{2}\,t_{ff}.
\end{eqnarray}

\ni This NE modification of the FFT leads to a new FFT $t^q_{ff}$ that relies on the nonextensive $q$-parameter as follows:
\vskip .2cm
\noindent (i) if $q=1 \Rightarrow t^q_{ff}=t_{ff}$.
\vskip .1 cm
\noindent (ii) f $0<q<1$ the FFT is modified as $t^q_{ff}<t_{ff}$.
\vskip .1 cm
\noindent (iii) if $1<q<5/3$,  the FFT is modified as $t^q_{ff}>t_{ff}$.
\vskip .1 cm
\noindent (iv) if $q\rightarrow 5/3,\, t^q_{ff} \rightarrow \infty$, the self-gravitating system keeps stable.

Using Eq. (\ref{Gk}) in (\ref{ff}), we have derived the expression for the free fall time in the Kaniadakis statistics

\begin{eqnarray}
\label{ffk}
t^\kappa_{ff}=\left(\frac{(1+\frac{\kappa}{2})}{(1+\frac{3}{2}\kappa) \,2\kappa}\,
\frac{\Gamma{(\frac{1}{2\kappa}-\frac{3}{4})\,\Gamma{(\frac{1}{2\kappa}+\frac{1}{4})}}}{\Gamma{(\frac{1}{2\kappa}+\frac{3}{4})\,\Gamma{(\frac{1}{2\kappa}-\frac{1}{4})}}}\right)^\frac{1}{2} t_{ff}.
\end{eqnarray}

\ni This $\kappa$ modification of the FFT leads to a FFT $t^\kappa_{ff}$ that depends on the $\kappa$-parameter as follows:
\vskip .2cm
\noindent (i) if $\kappa=0 \Rightarrow t^\kappa_{ff}=t_{ff}$.
\vskip .1 cm
\noindent (ii) if $0<\kappa<2/3$,  the FFT is modified as $t^\kappa_{ff}>t_{ff}$.
\vskip .1 cm
\noindent (iii) if $\kappa\rightarrow 2/3,\,t^\kappa_{ff} \rightarrow  \infty$, the self-gravitating system is always stable.

In Figure 2 we have plotted $t^q_{ff}$, Eq. (\ref{ffq}), and $t^\kappa_{ff}$, Eq. (\ref{ffk}), both normalized by $t_{ff}$, Eq.(\ref{ff}), as function of $\kappa$. We have used relation (\ref{rel}).

\begin{figure}
	\begin{center}
		\label{t_q}
		\includegraphics[width=3.5in, height=3.5in]{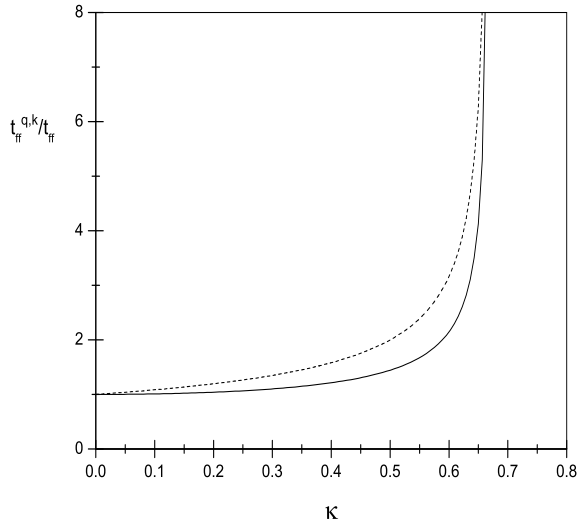}
		\caption{Dashed line: Tsallis critical time, $t_{ff}^q$. Solid line: Kaniadakis critical 
			time, $t_{ff}^\kappa$.}
	\end{center}
\end{figure}

From Figure 2 we can observe that, except for the limit values, $\kappa=0$ and $\kappa=2/3$, the Tsallis critical free fall time is greater than the Kaniadakis free fall time. This result indicates that the NE effects of Tsallis' statistics lead to a self-gravitating system to be more stable when compared with the effects produced by Kaniadakis' statistics since $t^q_{ff} > t^\kappa_{ff}$.


Before finishing this work, it is important to mention that many theoretical and experimental considerations\cite{mili1,mili2,mili3}, in a submillimeter range,  have suggested that the Newton law of gravitation can exhibit
a small deviation from inverse square law behavior. A recent paper \cite{new} shows that at separation down to $295 \mu m$, there are no deviations from Newton's law of gravitation.

The gravitational force would have the following form

\begin{eqnarray}
\label{mili}
F(r)=\frac{G m_1 m_2}{r^2} \, \left(1+\alpha \left(1+\frac{r}{\lambda}\right) e^{-r/\lambda}\right)\nonumber\\\nonumber\\=\frac{G(r) m_1 m_2}{r^2},
\end{eqnarray}
where $\alpha$ is a dimensionless parameter, $\lambda=\frac{\hbar}{m c}$ and the effective gravitational constant, $G(r)$, is given by

\begin{eqnarray}
\label{gef}
G(r)=G \left(1+\alpha \left(1+\frac{r}{\lambda}\right) e^{-r/\lambda}\right).
\end{eqnarray}

In the context of Verlinde's formalism, it is possible to rederive Eq.(\ref{gef}) by defining an effective Planck constant as

\begin{eqnarray}
\label{pcr}
\ell_P^2(r)=\ell_P^2\left(1+\alpha \left(1+\frac{r}{\lambda}\right) e^{-r/\lambda}\right),
\end{eqnarray}
and using  Eq.(\ref{pcr}) in (\ref{glp}) instead of $\ell_P^2$.

Comparing (\ref{gef}) when $r\approx 0$ with (\ref{S}),  we can find a relation for $\alpha$ and the nonextensive parameter $q$, given by

\begin{eqnarray}
\label{a1}
\alpha= \frac{3}{2}\,(1-q).
\end{eqnarray} 
When $q=1$ we have that $\alpha=0$. Comparing (\ref{gef}) when $r\approx 0$ with (\ref{Gk}), we can obtain a relation for $\alpha$ and the $\kappa$-parameter, written as

\begin{eqnarray}
\label{a2}
\alpha= \frac{(1+\frac{3}{2}\kappa) \,2\kappa}{(1+\frac{\kappa}{2})}\; \;
\frac{\Gamma{(\frac{1}{2\kappa}+\frac{3}{4})\,\Gamma{(\frac{1}{2\kappa}-\frac{1}{4})}}}{\Gamma{(\frac{1}{2\kappa}-\frac{3}{4})\,\Gamma{(\frac{1}{2\kappa}+\frac{1}{4})}} }-1.
\end{eqnarray} 
When $\kappa=0$ we have  that $\alpha=0$. Therefore,  it is possible to connect the parameter $\alpha$ of the modified gravitational force, Eq.(\ref{mili}), at a submillimeter range, with the Tsallis and  Kaniadakis parameters that are Eqs. (\ref{a1}) and (\ref{a2}).

To conclude,  in this work we have described both Tsallis and Kaniadakis nongaussian statistics in the light of Jeans mass criterion of stability. We have used the effective gravitational constants, Eqs. (\ref{S}) and (\ref{Gk}), in order to introduce Tsallis and Kaniadakis statistics in the self-gravitating systems. We believe that our procedure is simpler when compared to the methods previously used in the context of self-gravitating systems \cite{DJ}.
After some computations, the Tsallis Jeans mass and the Kaniadakis Jeans mass were compared trough comparing curves in Figure 1.  In Figure 2, the Tsallis critical free fall time and the Kaniadakis critical free fall time were shown.

\acknowledgments

\noindent The authors thank CNPq (Conselho Nacional de Desenvolvimento Cient\' ifico e Tecnol\'ogico), Brazilian scientific support federal agency, for partial financial support, Grants numbers 302155/2015-5, 302156/2015-1 and 442369/2014-0. E.M.C.A. thanks the hospitality of Theoretical Physics Department at Federal University of Rio de Janeiro (UFRJ), where part of this work was carried out.

\end{document}